\documentclass[pss,fleqn]{w-art}
\usepackage{times}
\usepackage{w-thm}
\usepackage{graphicx,bm,amsmath,amssymb,psfrag}

\newcommand{\mscr}[1]{{\mbox{\scriptsize #1}}}

\usepackage{graphicx,psfrag}
\begin{document}
%%    The information for the title page will be placed between
%%    \begin{document} and \maketitle. The order of most entries
%%    is determined by the class file and can not be changed by
%%    rearranging them. The maketitle command follows after the
%%    abstract.
%%
%%    Most of the following commands will be completed by the publisher.
%%
%%    The copyrightyear is defined in the .clo file as the first argument
%%    of the copyrightinfo command. If the copyrightyear differs from that
%%    value it might be adjusted by the following definition:
%%
%% \renewcommand{\copyrightyear}{2003}% uncomment to change the copyrightyear.
%%
\DOIsuffix{theDOIsuffix}
%%
%% issueinfo for header and copyright line
\Volume{XX}
\Issue{1}
\Month{01}
\Year{2003}
%%
%%    First and last pagenumber of the article. If the option
%%    'autolastpage' is set (default) the second argument may be left empty.
\pagespan{1}{}
%%
%%    Dates will be filled in by the publisher. The 'reviseddate' and
%%    'dateposted' (Published online) entry may be left empty.
%\Receiveddate{15 November 2003}
%\Reviseddate{30 November 2003}
%\Accepteddate{2 December 2003}
%\Dateposted{3 December 2003}
%%
\subjclass[pacs]{}

%% \pretitle{Editor's Choice}

%% We have a short and a long form for the title. The short form
%% (optional argument) goes into the running head.

\title[Short Title]{Spin densities in parabolic quantum wires with Rashba
  spin-orbit interaction}

%% Please do not enter footnotes or \inst{}-notes into the optional
%% argument of the author command. The optional argument will go into
%% the header.  If there is only one address the marker \inst{x} may be
%% omitted.

%% Information for the first author.
\author[Sh. First Author]{Sigurdur I.\ Erlingsson\footnote{Corresponding
     author: e-mail: {\sf sie@raunvis.hi.is}, Phone: +354\,525\,4708,
     Fax: +354\,525\,4708}\inst{1,2}} 
\address[\inst{1}]{Department of Physics
  and Astronomy, University of Basel, Klingelbergstrasse 82, CH-4056,
  Switzerland}
\address[\inst{2}]{Science Institute, University of Iceland, Dunhagi 3, IS-107
  Reykjavik, Iceland}
%%
%%    Information for the second author
\author[Sh. Second Author]{J.\ Carlos Egues
\inst{1,3}} 
\address[\inst{3}]{Department of Physics and Informatics, University of S\~{a}o Paulo at S\~{a}o Carlos, 13560-970
  S\~{a}o Carlos/SP, Brazil}
%%
%%    Information for the third author
\author[Sh. Third Author]{Daniel Loss\inst{1}}
%%
%%    \dedicatory{This is a dedicatory.}
\begin{abstract}
Using canonical transformations we diagonalize  approximately the Hamiltonian
of a gaussian wire with Rashba spin-orbit interaction.  This proceedure allows us to
obtain the energy 
dispersion relations and the wavefunctions with good accuracy, even in systems
with relatively strong Rashba coupling.  With these eigenstates one can
calculate the non-equilibrium spin densities induced by 
applying bias voltages across the sample.
We focus on the $z$-component of the spin density, which is related
to the spin Hall effect.
\end{abstract}
%% maketitle must follow the abstract.
\maketitle                   % Produces the title.

%% If there is not enough space inside the running head
%% for all authors including the title you may provide
%% the leftmark in one of the following three forms:

%% \renewcommand{\leftmark}
%% {First Author: A Short Title}

%% \renewcommand{\leftmark}
%% {First Author and Second Author: A Short Title}

%% \renewcommand{\leftmark}
%% {First Author et al.: A Short Title}

%% \tableofcontents  % Produces the table of contents.

The spin-orbit
interaction opens up the possibility to manipulate the electron spin using
electrical means, either with applied bias or gate voltage \cite{awschalom02:xx}. 
One manifestation of such electrical spin control is the spin Hall effect
\cite{sinova04:126603,chalaev05:245318,schliemann06:1015} where spin currents
are created via an interplay between spin-orbit interaction and the applied
electric field. 
When the spin Hall effect is considered in systems with finite transverse size
the edges plays an important role.  In disordered systems one would expect a
building up of spin density (spin accumulation) with opposite sign at the two edges.  
This picture is not necessarily valid in ballistic systems and the
wavefunctions (or wavepackets) themselves become important quantities
\cite{usaj05:631}.  The wavefunctions can then be used to
calculate the bias voltage induced spin density
\cite{lee05:045353,reynoso06:115342,wang06:033316,yao06:033314,bellucci06:045329,debald05:115322,mirales01:024426,serra05:235309,governale02:073311,nikolic05:075335}. 

In this paper we calculate analytically an approximate eigenspectrum of the Rashba
Hamiltonian in a gaussian quantum wire.  This allows us to calculate analytically the
relevant matrix elements and density of states for each transverse mode in the
wire with good accuracy, even for relatively strong spin-orbit
coupling. 
The Hamiltonian for the quantum wire made in a 2DEG with Rashba
spin-orbit interaction, for a given wavevector $k$ along the wire, is
\begin{eqnarray}
H&=&\frac{\hbar^2 k^2}{2m}+\frac{1}{2} m \omega^2 y^2
+\frac{\alpha}{\hbar}(\hbar k\sigma_y-p_y\sigma_x) \\
&=&\frac{\hbar^2 k^2}{2m}
+\hbar \omega (a^\dagger a +\frac{1}{2})
+ \tau_z\alpha k
-\frac{\alpha}{\sqrt{2} \ell} (a^\dagger\tau_+ +a\tau_-) 
+\frac{\alpha}{\sqrt{2} \ell} (a^\dagger\tau_- +a\tau_+) .
\label{eq:Hamiltonian}
\end{eqnarray}
In Eq.\ (\ref{eq:Hamiltonian}) we introduced for convenience new spin matrices
where 
$\tau_z=\sigma_y$ and $\tau_\pm$ raises and lowers between the $\sigma_y$
eigenstates.  For positive momenta only states $|k,n,\uparrow \rangle$ and
$|k,n+1,\downarrow \rangle$ cross in energy.  All other states are
separated by at 
least $\hbar\omega$, which allows us to treat the term
proportional to $(a^\dagger\tau_+ +a\tau_-)$ perturbatively in
$k_\mscr{so}\ell\equiv 
\alpha/\ell \hbar \omega$.
Note that our Hamiltonian is time reversal symmetric (there is no external
source of magnetic field).  Hence we can focus on
the posive $k$ states and use the Krames relations
$\epsilon_{n,s}(k)=\epsilon_{n,-s}(-k)$ and
$\psi_{k,n,s}(\bm{r})=\psi_{-k,n,-s}(\bm{r})$ to obtain the negative $k$
states.  

\begin{figure}[t]
\begin{center}
\psfrag{Energy}{\Huge $\varepsilon_{ns}(k)/\hbar \omega$}
\psfrag{figA}{\Huge $(a)$}
\psfrag{figB}{\Huge $(b)$}
\psfrag{kell}{\Huge $k \ell $}
\includegraphics[angle=-90,width=7.0cm]{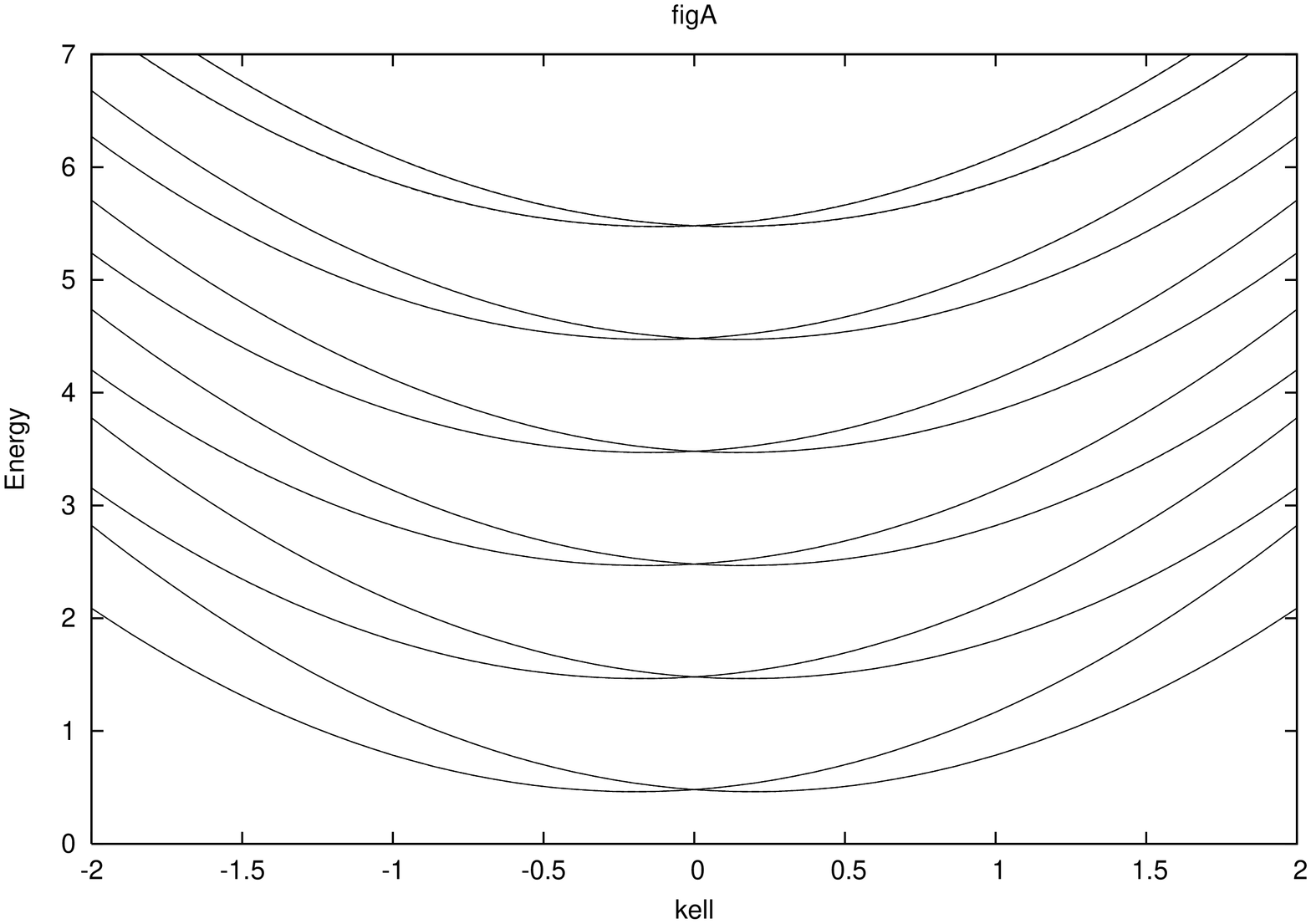}
\includegraphics[angle=-90,width=7.0cm]{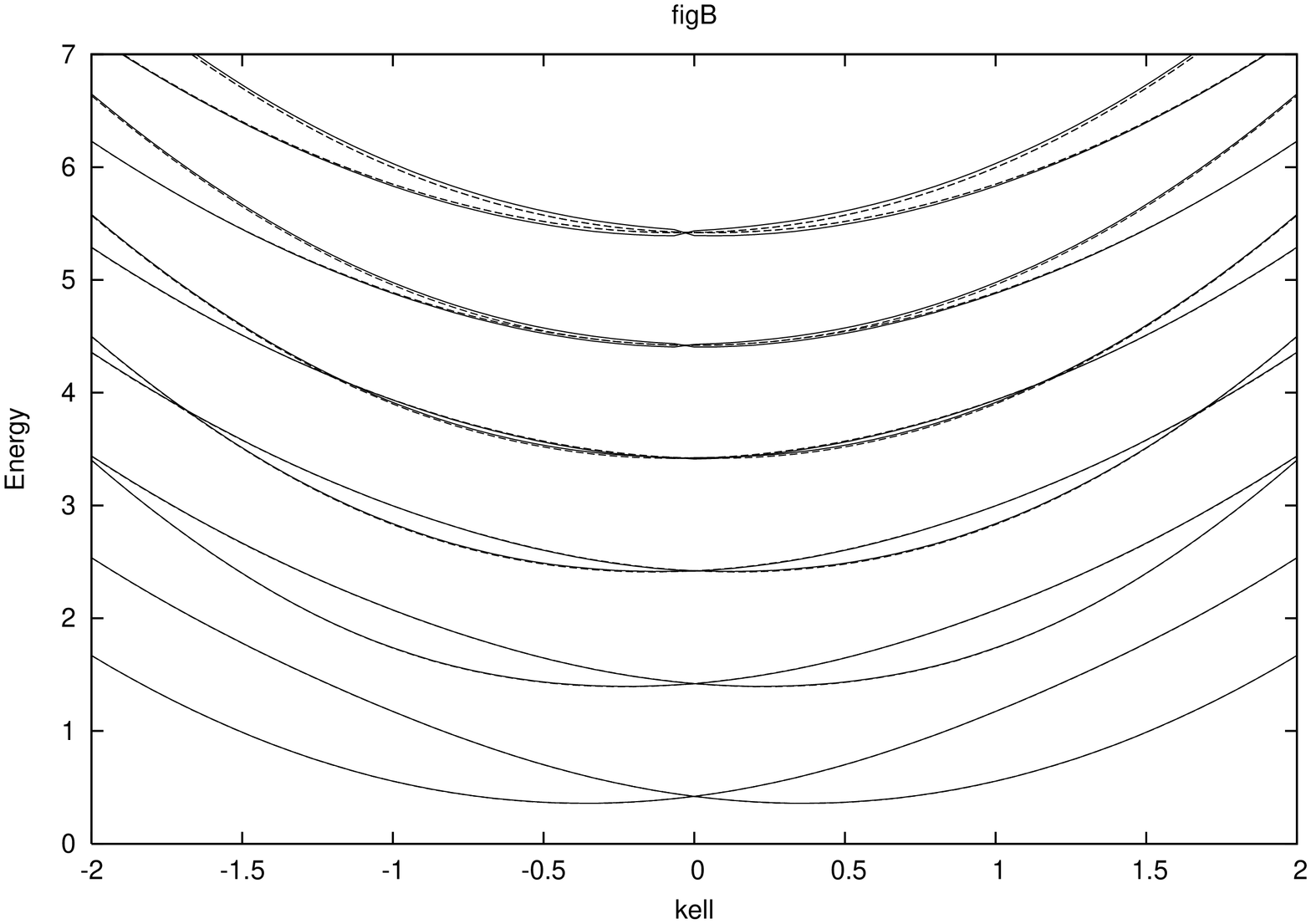}
\caption{Energy dispersions of Eqs.\ (\ref{eq:E0D})-(\ref{eq:EnU}) are plotted
  (solid lines) for the first 6 oscillator levels for 
  $k_\mscr{so}\ell=0.2$ in $(a)$ and $k_\mscr{so}\ell=0.4$ in $(b)$.  The
  dashed lines are the results of numerical calculations.  Note that the
  numerical curves fall ontop of the analytical curves except for the highest
  energy curves in $(b)$.}
\label{fig:dispersion}
\end{center}
\end{figure}

As mentioned above, for the positive momentum states the perturbative term is 
\begin{eqnarray}
V&=&-\frac{\alpha }{2\sqrt{2} \ell} (\tau_+a^\dagger+ \tau_- a),
\end{eqnarray}
and the non-perturbed Hamiltonian is 
\begin{eqnarray}
H_0&=&\frac{\hbar^2k^2}{2m}
+\hbar \omega (a^\dagger a +\frac{1}{2})
+ \alpha \tau_z k
+ \frac{\alpha}{\sqrt{2} \ell} (\tau_+ a+\tau_ - a^\dagger).
\end{eqnarray}
We define an effective Hamiltonian $H_\mscr{eff}=e^S He^{-S}$ which results in 
\begin{eqnarray}
H_\mscr{eff}&=&H_0+\frac{1}{2}[S,V]+\frac{1}{3!}[S,[S,V]]+o((k_\mscr{so}\ell)^4),
\label{eq:HeffExpansion}
\end{eqnarray}
with $S$ chosen to fulfill the usual condition $V+[S,H_0]=0$.  The $S$ which
satisfies the above requirement is
\begin{eqnarray}
S&=&\frac{\alpha }{2\sqrt{2} \ell(\hbar \omega+2\alpha k)} (\tau_+a^\dagger- 
\tau_- a)
+\frac{\alpha^2 }{4 \ell^2\hbar \omega (\hbar \omega+2\alpha k)} 
\tau_z(
{a^\dagger}^2-{a}^2) +o((k_\mscr{so}\ell)^3).
\end{eqnarray}
The new terms generated by $S$ in the effective Hamiltonian are
\begin{eqnarray}
\frac{1}{2}[S,V]&=&
\frac{\alpha^2 (\tau_z(2 a^\dagger a +1)-1)}{4\ell^2(\hbar \omega+2\alpha k)}
+\frac{\alpha^3 \bigl (
\tau_+a a^\dagger a +\tau_+ {a^\dagger}^3+\mbox{h.c.}
\bigr )}{\sqrt{2} 4 \ell^3\hbar \omega(\hbar \omega+2\alpha k)} 
+o((k_\mscr{so}\ell)^4).
\label{eq:commutatorSV}
\end{eqnarray}
The term proportional to $\tau_+{a^\dagger}^3$ and its hermitian conjugate
are perturbative, i.e.\ they only couple terms which are separated 
by at least $3\hbar \omega$.  Furthermore, one can show that the term
$\frac{1}{3}[S,[S,V]]$ in Eq.\ (\ref{eq:HeffExpansion}) contains no 'diagonal'
terms\footnote{Here,
  diagonal refers to $a^\dagger a$, $\tau_z$ and also
  the terms containing $\tau_+ a$ and $\tau_- a^\dagger$.}
and thus the corrections 
to the Hamiltonian, and consequently its eigenenergies, are at most
$o((k_\mscr{so}\ell)^4)$. 
The resulting effective Hamiltonian is thus
\begin{eqnarray}
H_\mscr{eff}&=&
\frac{\hbar^2k^2}{2m}
+
\hbar \omega (a^\dagger a+1/2)
+
\alpha k \tau_z +\frac{\alpha^2(\tau_z(2a^\dagger a +1)-1)}{4\ell (\hbar
  \omega+2 \alpha k)}
\nonumber \\
& &
-\frac{\alpha}{2\sqrt{2}\ell}
\left (
\tau_+ a  \left (
1-\frac{\alpha^2a^\dagger a }{4 \ell \hbar \omega (\hbar \omega +2\alpha
  k)}  \right  )
+
\left (
1-\frac{\alpha^2a^\dagger a }{4 \ell \hbar \omega (\hbar \omega +2\alpha
  k)}  \right  )
\tau_- a^\dagger 
\right ).
\label{eq:Heff}
\end{eqnarray}
The above $H_\mscr{eff}$ 
can be exactly diagonalized using a
generalized rotation matrix in spin space
\begin{eqnarray}
U&=&
\left (
\begin{array}{cc}
\cos (\Theta[\hat{n}+1]/2) & \frac{\sin
  (\Theta[\hat{n}+1]/2)}{\sqrt{\hat{n}+1}} a \\ 
-a^\dagger \frac{\sin
  (\Theta[\hat{n}+1]/2)}{\sqrt{\hat{n}+1}}& 
\cos (\Theta[\hat{n}]/2)
\end{array} \right ) ,
\label{eq:U}
\end{eqnarray}
where 
$\hat{n}=a^\dagger a$ and 
\begin{eqnarray}
\cos(\Theta[\hat{n}+1])=\frac{\left ( \frac{1}{2}- k_\mscr{so}\ell^2
    k-\frac{(k_\mscr{so}\ell)^2 (\hat{n}+1)}{2 (1+2 k_\mscr{so}\ell^2 k)} \right )}{\sqrt{\left ( \frac{1}{2}- k_\mscr{so}\ell^2
    k-\frac{(k_\mscr{so}\ell)^2 (\hat{n}+1)}{2 (1+2 k_\mscr{so}\ell^2 k)} \right )^2
+ \frac{(k_\mscr{so}\ell)^2 (\hat{n}+1)}{2} \left ( 1-\frac{(k_\mscr{so}\ell)^2 (\hat{n}+1)}{4(1+2
    k_\mscr{so}\ell^2 k)}\right )^2}}.
\label{eq:cos}
\end{eqnarray}
The resulting Hamiltonian $H_\mscr{diag}=U^\dagger
H_\mscr{eff}U$ is diagonal in the spin and ladder operators and its eigenenergies
for positive $k$ are
\begin{eqnarray}
\varepsilon_{0,\downarrow}(k)&=&\hbar \omega \left (
\frac{1}{2}\ell^2k^2+\frac{1}{2}-k_\mscr{so}\ell^2 k
-\frac{(k_\mscr{so}\ell)^2}{2(1+2k_\mscr{so}\ell^2 k)} 
\right ) \label{eq:E0D}\\
\varepsilon_{n,\downarrow}(k)&=& \hbar \omega \left (
\frac{1}{2}\ell^2 k^2+n-\frac{(k_\mscr{so}\ell)^2}{2(1+2k_\mscr{so}\ell^2 k)}
+\Delta_{n}(k)
\right ) ,\quad n>0 \label{eq:EnD}\\
\varepsilon_{n,\uparrow}(k)&=& \hbar \omega \left (
\frac{1}{2}\ell^2 k^2+(n+1)-\frac{(k_\mscr{so}\ell)^2}{2(1+2k_\mscr{so}\ell^2 k)}
-\Delta_{n+1}(k) \right )
,\quad n \geq 0 \label{eq:EnU} \\
\Delta_n(k)&=&\sqrt{\left ( \frac{1}{2}- k_\mscr{so}\ell^2
    k-\frac{(k_\mscr{so}\ell)^2 n}{2 (1+2 k_\mscr{so}\ell^2 k)} \right )^2
+ \frac{(k_\mscr{so}\ell)^2}{2} n\left ( 1-\frac{(k_\mscr{so}\ell)^2 n}{4(1+2
    k_\mscr{so}\ell^2 k)}\right )^2}.
\end{eqnarray}
The negative $k$ branch of the spectrum is obtained from the Kramers relations.
In Fig.\ \ref{fig:dispersion} the energy dispersion for the full spectrum is
plotted, along with results of numerical calculations (dashed curves).  
Even for a relatively large ($k_\mscr{so}\ell=0.4$) perturbation parameter the
numerical and 
analytical curves are barely distinguishable except for the highest transverse
states, see Fig.\ \ref{fig:dispersion}b.

{\it Spin densities: } 
Knowing the wavefunctions and eigenenergies allows one to calculate many
physical quantities analytically, even non-equilibrium ones.  As an example we
consider spin densities (transverse to the wire direction) and focus on the
the average value of $\sigma_z\delta(y)$.  For an applied bias voltage,
$\langle \sigma_z \delta(y)\rangle$ is related to the spin-Hall
effect \cite{nikolic05:046601}. 
Since the Hamiltonian is time reversal symmetric there will be no equilibrium
spin density in the system \cite{rashba03:241315R}.  Applying a bias voltage
lifts this symmetry since Kramers pairs become unequally occupied.  
For spin densities (and any quantity which is odd under time reversal) 
only states in the bias window need to be considered.  Starting from the
non-equilibrium Green's functions\cite{datta95:xx} it is possible to show that for
$kT=0$ the spin density is 
\begin{eqnarray}
\langle \tau_\eta \delta(y) \rangle|_{eV}&=&
\sum_{kns} \langle \psi_{kns}|\tau_\eta \delta(y) |\psi_{kns}\rangle
\Bigl [ f \bigl (\varepsilon_{ns}(k)-(\mu+eV)\bigr ) -f\bigl ( \varepsilon_{ns}(k)-\mu
\bigr ) \Bigr ],
\label{eq:densitySeta}
\end{eqnarray}
where the sum is
over all $k$ whose velocity is
positive (assuming that $eV>0$), i.e.\ the sum contains not only $k>0$
states but can also include negative momentum states.  
Although the bias induced spin densities are know analytically through the
wavefunctions, the equations are quite long and complicated.  As an example we
give the following matrix element
\begin{eqnarray}
\langle \psi_{kn\uparrow}|\tau_x\delta(y)| \psi_{kn\uparrow} \rangle &=&
-\sin (\Theta[n+1])\phi_{n}(y)\phi_{n+1}(y) \nonumber \\
& +& \frac{ k_\mscr{so}\ell}{1+2k_\mscr{so}\ell^2 k}
\Bigl \{ \cos (\Theta[n+1]/2)^2\phi_{n}(y)\phi_{n-1}(y) \nonumber \\ 
& &\quad \quad -\sin (\Theta[n+1])/2)^2\phi_{n+1}(y)\phi_{n+2}(y) \Bigr \},
\label{eq:matrixElement}
\end{eqnarray}
where $\langle y| n \rangle=\phi_n(y)$ are the harmonic oscillator
eigenstates. 
The matrix element in Eq.\ (\ref{eq:matrixElement}) has the
symmetry property $\langle \psi_{kns}|\sigma_z\delta(y)|
\psi_{kns} \rangle =-\langle \psi_{kns}|\sigma_z\delta(-y)|
\psi_{kns} \rangle$.  This is consistent with the spin-Hall effect
phenomenology where the transverse spin current leads to opposite spin
polarization at the two edges \cite{nikolic05:046601,usaj05:631}.  This
property is a direct result of the parity of the harmonic oscillator states
$\phi_n(y)$.  

\begin{figure}[t]
\begin{center}
\psfrag{y-axis}{\Huge $y/\ell$}
\psfrag{figA}{\Huge $(a)$}
\psfrag{figB}{\Huge $(b)$}
\psfrag{kSOl0.30}[r]{\Huge $k_{so}\ell=0.3$}
\psfrag{densitySz}{\Huge $\langle \sigma_z \delta(y)\rangle \left [ \frac{
      eV}{\hbar \omega \ell^2}\right ]  $}
\psfrag{line1}[r]{\Large $\mu=1.5$}
\psfrag{line2}[r]{\Large $\mu=1.7$}
\psfrag{line3}[r]{\Large $\mu=1.9$}
\psfrag{line4}[r]{\Large $\mu=2.1$}
\psfrag{line5}[r]{\Large $\mu=2.3$}
\includegraphics[angle=-90,width=7.0cm]{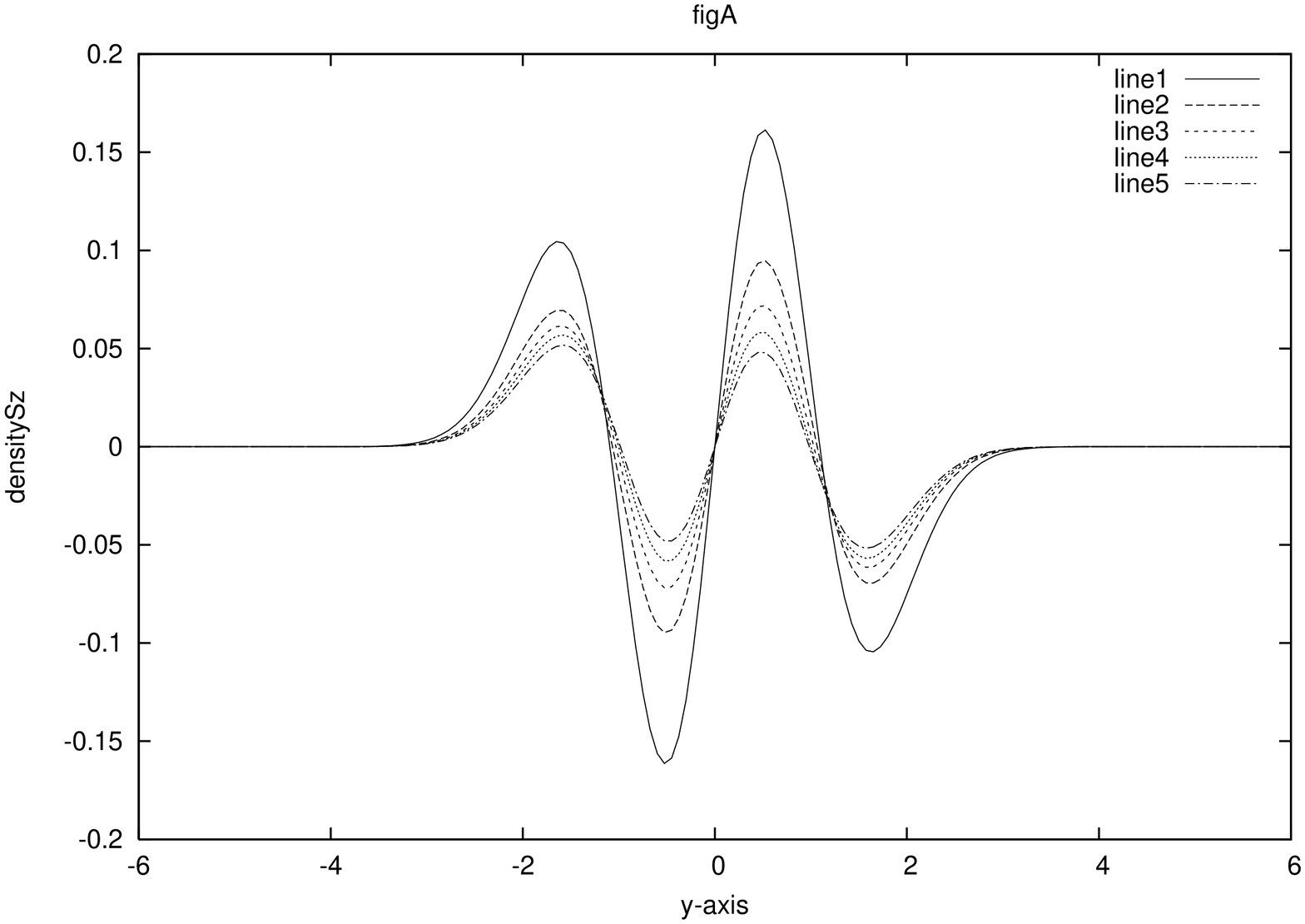}
\psfrag{line1}[r]{\Large $\mu=3.5$}
\psfrag{line2}[r]{\Large $\mu=5.7$}
\psfrag{line3}[r]{\Large $\mu=3.9$}
\psfrag{line4}[r]{\Large $\mu=4.1$}
\psfrag{line5}[r]{\Large $\mu=4.3$}
\includegraphics[angle=-90,width=7.0cm]{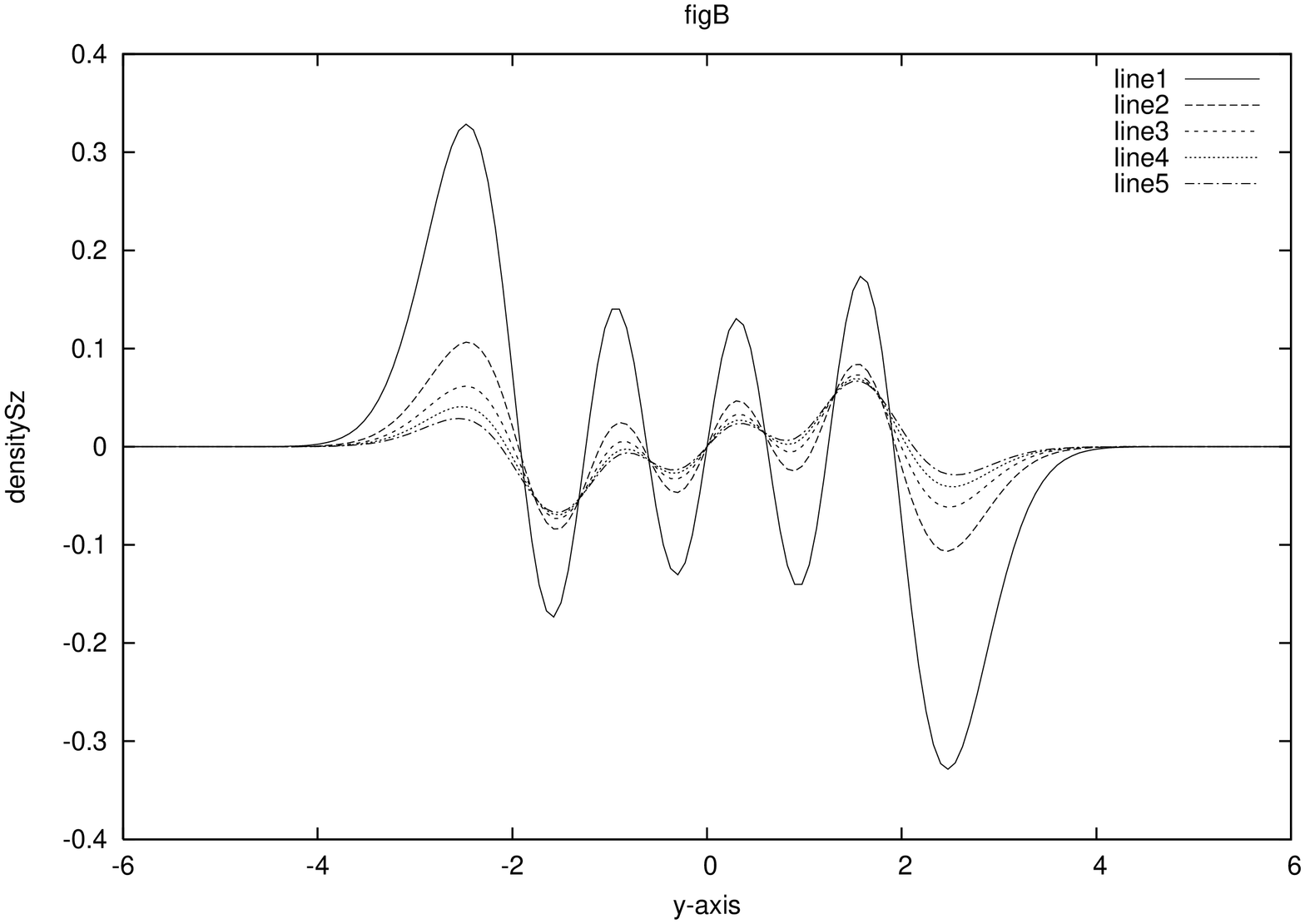}
\caption{The spin-density for different $\mu$  such that 2 $(a)$, and 4
  $(b)$ harmonic oscillator levels for $k_\mscr{so}\ell=0.4$.  The
  spin-density is largest close to each 
  band bottom, where the density of states is greatest.} 
\label{fig:densitySz}
\end{center}
\end{figure}

The spin densities (normalized to the wire length) for different values of
the chemical potential $\mu$ are plotted in Fig.\ \ref{fig:densitySz},
assuming 
linear response $eV \ll \hbar \omega$.  As expected the spin
density is odd in $y$ and the highest transverse mode contributes the most
due to its density of states being the highest. Note that the size of the spin
density is comparable to the induced particle density since the matrix element
in Eq.\ (\ref{eq:densitySeta}) is order unity due to the strong coupling of
adjacent oscillator states with opposite spin (see Eq.\ (\ref{eq:Heff}))

To summarize, we have calculated the eigenspectrum of a gaussian quantum wire
with Rashba interaction.  The spectrum allows us to calculate the
non-equilibrium spin density analytically.  
We presented results for the $\sigma_z$ density for a few low lying oscillator
states but the formalism can well handle higher states also.  
\begin{acknowledgement}
This work was supported by the Icelandic Research Fund, the Swiss NSF, the NCCR Nanoscience, EU NoE MAGMANet,  DARPA, ARO, ONR, JST ICORP, CNPq, and FAPESP. 
\end{acknowledgement}

\end{document}